# Origin of room temperature ferromagnetism in homogeneous (In,Mn)As thin films

A. J. Blattner, P. L. Prabhumirashi, V. P. Dravid, B. W. Wessels[*]

*Department of Materials Science & Engineering & Materials Research Center, Northwestern University, Evanston IL 60208*

**Abstract**

The microstructure of (In,Mn)As thin films grown using metalorganic vapor phase epitaxy (MOVPE) was investigated to determine the origin of room temperature ferromagnetism in these films. Transmission electron microscopy (TEM) based techniques were used to investigate phase purity and compositional homogeneity. Microanalysis of an $In_{1-x}Mn_xAs$ film with $x = 0.01$ and a Curie temperature of 330 K exhibited a homogeneous distribution of Mn. High Mn concentration films with $x = 0.20$ exhibited MnAs precipitates within the (In,Mn)As matrix. The analysis indicates that room temperature ferromagnetic, single-phase (In,Mn)As can be formed by MOVPE. The origin of ferromagnetism is attributed to (In,Mn)As solid solution rather than distinct secondary Mn-rich magnetic phase(s).

---

[*] Corresponding author. Tel.: +1-847-491-3219; fax: +1-847-491-7820;
*E-mail address*: b-wessels@northwestern.edu (B. W. Wessels).

The diluted magnetic semiconductor (DMS) (In,Mn)As holds great promise for future spintronics applications. The combination of its magnetic and semiconducting properties opens up exciting possibilities for novel devices. (In,Mn)As is the only DMS to exhibit light induced ferromagnetism [1,2] as well as electric-field control of ferromagnetism [3]. We have reported on room temperature ferromagnetism in In$_{1-x}$Mn$_x$As layers with $x$ = 0.01-0.10 grown using metalorganic vapor phase epitaxy (MOVPE) [4]. Single-phase films, as determined by x-ray diffraction, exhibited a Curie temperature of 333 K, in contrast to Curie temperatures of 35-100 K observed in thin films grown by molecular beam epitaxy (MBE) [5,6]. Room temperature ferromagnetism in MBE grown material has been previously attributed to the presence of ferromagnetic MnAs precipitates which have a Curie temperature of approximately 315 K [7].

As to MOVPE grown films, although x-ray scattering shows lack of secondary magnetic phases, the detection limit is often about five volume percent and isostructural secondary phases are not easily accounted for. Thus, to further probe the origin of room temperature ferromagnetism in the films grown by MOVPE, we have performed a microstructural evaluation using transmission electron microscopy (TEM). Plan view samples indicated that a film exhibiting room temperature ferromagnetism was phase pure with no detectable presence of MnAs. Conversely, other films with higher Mn content exhibited MnAs nanoprecipitates within the (In,Mn)As matrix. Energy dispersive spectrometry (EDS) was used to examine the Mn distribution within these films.

(In,Mn)As films were deposited using atmospheric pressure MOVPE on semi-insulating GaAs(001) substrates. Detailed growth procedures have been previously described [8]. Plan-view TEM specimens were prepared using a chemical thinning procedure previously used for heteroepitaxial films on GaAs substrates [9]. Using this technique, it was possible to expose large areas of (In,Mn)As film for imaging with minimal damage associated with ion milling. Cross sectional TEM specimens were prepared using conventional dimpling and ion milling procedures. TEM analysis was performed on a Hitachi HF-2000 atomic resolution analytical electron microscope, equipped with a cold-field emission gun, which is operable at accelerating voltages up to 200 keV. EDS measurements were done using a 2 nm (at full width tenth maximum) probe, giving a spatial resolution of 2.8 nm. The details regarding the spatial resolution calculations, using Monte Carlo techniques, have been published elsewhere [10].

For these studies, both single-phase ($x = 0.01$) and two-phase ($x = 0.20$) samples were deposited using MOVPE at 500 °C and 440 °C, respectively. The nominal Mn concentrations were determined using EDS measurements. A superconducting quantum interference device (SQUID) magnetometer (Quantum Design, model MPMS) was used to measure the temperature dependent magnetization from 5-350 K. After subtraction of the diamagnetic contribution of the GaAs substrate, the resulting magnetization of the single-phase $In_{0.99}Mn_{0.01}As$ sample is shown in Fig. 1. The Curie temperature for this film was approximately 330 K. A cross sectional TEM image is shown in Fig. 2(a), indicating structurally homogeneous film formation.

A plan view TEM specimen was prepared from the In$_{0.99}$Mn$_{0.01}$As film. Scanning transmission electron microscopy (STEM) was used to observe mass-thickness contrast differences within the TEM specimen. Figure 2(b) (inset) is the dark-field STEM image taken from the In$_{0.99}$Mn$_{0.01}$As specimen. Other than some inevitable strain contrast from dislocations in the film, no obvious mass-thickness contrast could be observed. A large portion of the film (> 1 μm$^2$) was examined for contrast changes due to the presence of MnAs precipitates, but resulted in images similar to Fig. 2(b), providing indirect evidence for chemical homogeneity in the film. An EDS line profile scan was performed to determine the compositional homogeneity of this sample. The integrated intensity ratios for In (L$_\alpha$ and L$_\beta$) and Mn (K$_\alpha$), as normalized to As (K$_\alpha$), are shown in Fig. 2(b). The inset image indicates where the line scan was taken. Absence of any pronounced mass-thickness contrast in the annular dark field STEM image as well as constant In/As and Mn/As EDS integrated intensity ratios indicates the sample is homogeneous and lacks the presence of any nanoprecipitates.

A TEM specimen of a multi-phase (In,Mn)As sample with a nominal composition of $x$ = 0.20 was investigated. MnAs as a second phase was previously observed in this sample by x-ray diffraction measurements. Figure 3(a) is the bright field TEM image of this multi-phase sample. Ellipsoidal MnAs precipitates ranging from 100-400 nm in size are clearly visible and are uniformly distributed throughout the matrix. Figure 3(b) shows the selected area diffraction pattern (SADP) along the (In,Mn)As[001] zone axis. One can see the double-diffraction spots surrounding primary (In,Mn)As reflections. Reflections attributed to double diffraction are a common feature of SADPs recorded from two-phase materials. In this case, the double-diffraction spots manifest the presence of MnAs as a second phase. An EDS line scan across a precipitate, as seen in Fig. 4 shows an increase in the Mn/As integrated intensity ratio with a corresponding proportional decrease in the In/As ratio.

The preceding analysis demonstrates the ability to resolve MnAs nanoprecipitates within an (In,Mn)As matrix using TEM and STEM techniques. The absence of MnAs nanoprecipitates in the $In_{0.99}Mn_{0.01}As$ film supports the assertion that it is truly single-phase and that the room temperature ferromagnetism is not a result of distinct MnAs secondary phases.

The high Curie temperature observed for this single-phase film suggests that the Mn has a different site distribution than in films grown by low temperature MBE, which have $T_c$ below 60 K. Recently we have proposed a model for this ferromagnetism within the framework of atomic cluster formation, whereby the Mn is present in the form of dimers and trimers located on cation sites [11]. This model is more appropriate for the present case where the films are grown at temperatures 200 degrees higher than are typically used for MBE grown III-V DMS. The increased incorporation kinetics at these higher growth temperatures allow for the formation of Mn atomic clusters that stabilizes the ferromagnetism. Extended x-ray absorption fine structure (EXAFS) and electron energy loss spectroscopy (EELS) studies are needed to address the atomic distribution of Mn within the (In,Mn)As films and are presently underway.

In conclusion, we have determined the microstructure of single-phase and multi-phase (In,Mn)As films grown by MOVPE. The low Mn concentration film, which exhibited room temperature ferromagnetism, is single-phase as determined by TEM and does not exhibit MnAs nanoprecipitate formation. In contrast, the high Mn concentration film did exhibit

MnAs nanoprecipitate formation, which was easily observed in both TEM and STEM imaging modes. These MnAs precipitates range in size from 100-400 nm. The present work indicates that room temperature ferromagnetic thin films can be prepared using this materials system, and points to atomic-level origin of ferromagnetism in the single-phase (In,Mn)As alloys, rather than distributed secondary phases.

This work was supported, in part, by the National Science Foundation through the MRSEC program under grant number DMR-0076097 and the Spin Electronics Program under ECS-0224210. The authors wish to acknowledge extensive use of the Materials Research Center Facilities at Northwestern University.

**Figure Captions**

Fig. 1. Temperature dependent magnetization at 10 kOe applied magnetic field for a single-phase $In_{0.99}Mn_{0.01}As$ film. The magnetic field was applied perpendicular to the plane of the film and the sample was zero-field cooled.

Fig. 2. (a) TEM image of cross sectional $In_{0.99}Mn_{0.01}As$. Also shown is a high-resolution TEM image with (In,Mn)As {111} lattice planes resolved. MnAs precipitates were not observed. The film had a Curie temperature of 330 K. (b) Line profile across $In_{0.99}Mn_{0.01}As$/GaAs(001) plan view sample in which the EDS integrated intensity ratios (In/As and Mn/As) are plotted as a function of distance after Ref. 12. The region of the line scan is indicated in the inset annular dark field STEM image, while the box indicates the area used for auto-beam drift correction. The dark regions in the inset STEM image are regions that were over-etched during TEM specimen preparation.

Fig. 3. (a) Bright field image of MnAs precipitates distributed in the $In_{1-x}Mn_xAs$ matrix. Precipitates were 100-400 nm in size. (b) Selected area diffraction pattern from (In,Mn)As [001] zone axis showing double-diffraction spots, occurring due to the presence of MnAs second phase, around primary (In,Mn)As reflections.

Fig. 4. Line profile across a MnAs precipitate in $In_{1-x}Mn_xAs$ + MnAs in which EDS integrated intensity ratios (In/As and Mn/As) are plotted as a function of distance after Ref. 12.

**FIG. 1—Blattner et al.**

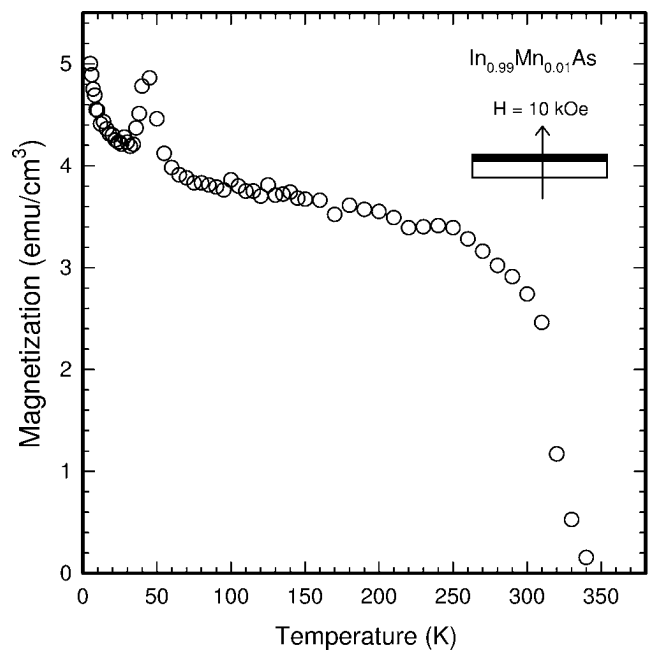

# FIG. 2—Blattner et al.

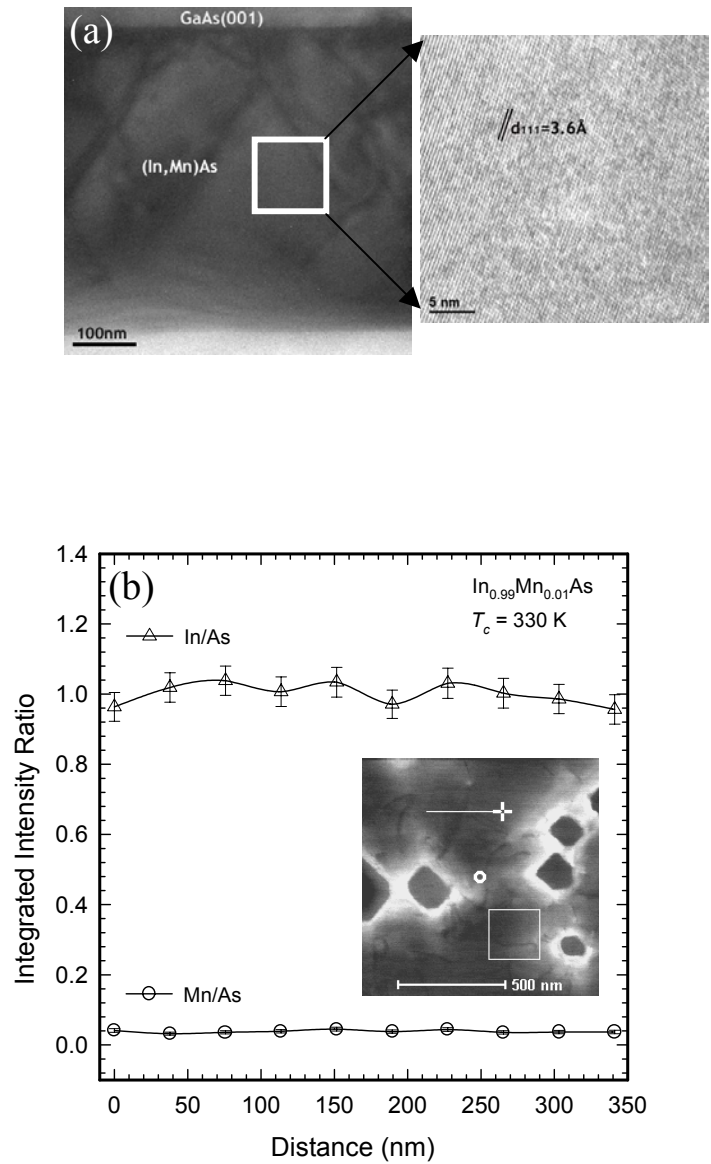

# FIG. 3—Blattner et al.

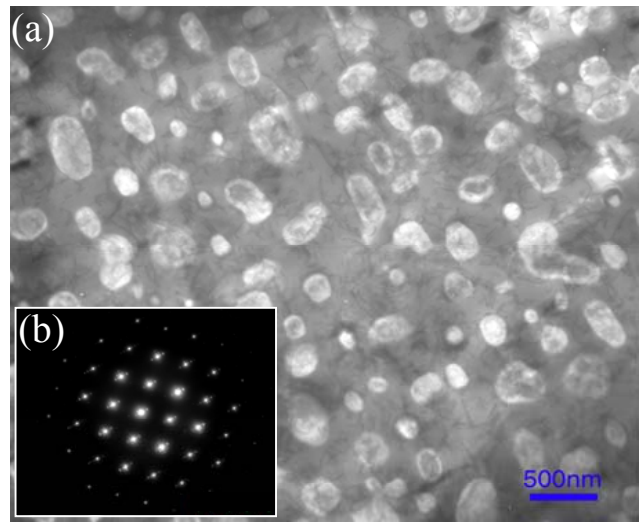

**FIG. 4—Blattner et al.**

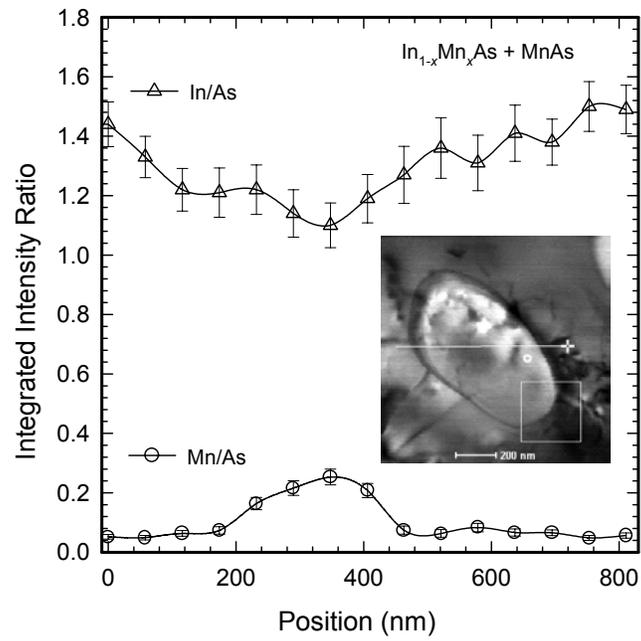